\documentclass[12pt,prd,aps,floats,superscriptaddress,nofootinbib,preprint]{revtex4}
\usepackage{graphicx}
\usepackage{amssymb}

\newcommand{\beq}[1] {\begin{equation}\label{#1} }
\newcommand{\eeq} {\end{equation} }

\newcommand{\bea}[1]{\begin{eqnarray}\label{#1} }
\newcommand{\eea}{\end{eqnarray}}

\newcommand{\si}{\sigma}
\newcommand{\eps}{\epsilon}
\newcommand{\te}{\theta}
\newcommand{\del}{\delta}

\begin{document}

\title{Precision electroweak constraints on 
Universal Extra Dimensions revisited}

\author{Ilia Gogoladze \footnote{On a leave of absence from:
Andronikashvili Institute of Physics, GAS, 380077 Tbilisi,
Georgia.}} \email{ilia@physics.udel.edu} \affiliation{\it Department
of Physics and Astronomy, University of Delaware, Newark, DE 19716,
USA }

\author{Cosmin Macesanu}
\email{cmacesan@physics.syr.edu}
\affiliation{\it Department of Physics, Syracuse University
Syracuse, New York 13244}

\begin{abstract}
We reconsider the constraints on Universal Extra Dimensions (UED) models
arising from precision electroweak data. We take into account 
the subleading contributions from new physics (expressed in
terms of the $X,Y\ldots$ variables), as well as 
two loop corrections to the Standard Model $\rho$ parameter.
For the case of one extra dimension, we obtain a lower bound on the inverse
compactification scale $ M = R^{-1}$
of 600 GeV (at 90\% confidence level),  with a Higgs mass of 115 GeV.
However, in contradiction to recent claims, we find that 
this constraint is significantly relaxed  with increasing Higgs mass,
allowing for compactification scales as low as 300 GeV.
LEP II data does not affect significantly these results.

\end{abstract}

\maketitle

\section{Introduction}
\label{sec:intro}

Theories where the Standard Model fields propagate in
large extra dimensions may lead to testable predictions
for the direct observation of new particles (the
Kaluza Klein excitations of the SM fields)
 at high energy colliders. However,  these
new particles may already make their presence known through loop corrections
to low energy observables.

The predictions of the Standard Model agree extremely well with
the wealth of data accumulated over the years in  collider experiments at
energies of order 100 GeV and below. This agreement leads to tight
constraints on
any new physics.  One of the more important set of constraints
is obtained by the measurements of fundamental parameters of the electroweak
theory (like the gauge bosons masses, the $Z$ boson decay partial and total
decay widths, or the effective Weinberg mixing angle) by the experiments
taking place at the Large Electron Positron (LEP) collider. In particular,
for models where only
gauge bosons propagate in the bulk \cite{Masip-mix}, the LEP data
requires  that the masses of the new KK excitations be so large
(a few TeV) that is impossible to see them at the Tevatron,
and one has to wait for the Large Hadron Collider.

 However, for a particular class of models, where all the SM fields
propagate in extra dimensions \cite{acd},
the precision constraints are relatively weak. Such universal extra
dimensions (UED) models
enjoy the property of KK number conservation
(which arises as a consequence of conservation of momentum in the extra
dimensions). As a consequence, KK excitations must be produced in pairs
(at least at tree level), and their contribution to processes taking place
at energies below 100 GeV  is relatively suppressed.

The strongest constraints on the compactification scale of UED models still
comes from observables measured at the $Z$ pole, in particular, the relation
between the $Z$ and $W$ gauge bosons masses.
Since a single KK excitation does not couple directly to the SM fermions,
the  heavy KK states only
contribute to the self energies of the gauge bosons, and their contribution
can be parametrized in terms of the $S,T,U$ \cite{Peskin_T} variables.
In particular, the constraint on the $T$ parameter is the more stringent
one, and taking it into account requires that the mass $M$ of the first level
KK excitations be higher than about 550 GeV \cite{a_yee}
(for a Higgs mass of order 100 GeV)\footnote{Note
that the original paper \cite{acd} has a factor
of 2 missing in the evaluation of $T$, and the constraint they cite is
significantly weaker: $M \gtrsim 350$ GeV. Also, the above result applies for 
the case of one extra dimension, which is the scenario discussed
in this paper.}. 
For comparision, constraints
from different physics (for example, flavor violating processes like
 $b\rightarrow s \gamma$ \cite{Agashe}, or the muon $g-2$ \cite{App-mu})
are of the order $M \gtrsim 280$ GeV or weaker. Also, an important feature
of the $Z$ pole constraints on the compactification scale $M$
is that it depends on the Higgs mass (as noted by \cite{a_yee}). It turns
out that the contributions of a heavy Higgs to the $T$ parameter have the
opposite sign to the leading constribution from KK states, and as such
increasing the Higgs mass relaxes the bound on $M$ considerably; for a Higgs
mass of order 500 GeV, one can have values of the compactification scale
as low as 300 GeV.

A more recent paper \cite{flacke} performs the analysis of the LEP data
constraints on $M$ by taking into account data taken above the $Z$ pole,
as well as the two loop electroweak corrections (involving the Higgs boson and
the top quark) to the $\Delta \rho$ parameter in the Standard Model.
The conclusion reached in this paper is quite striking: they find that
the constraint on the compactification scale
increases strongly with the Higgs mass (in contradiction to the result
derived in \cite{a_yee})
thus setting a lower limit on $M$ of about 800 GeV (at 95\% CL).
This has important
consequences for phenomenology, as well as for the cosmological implications
of the model. First, it would make the observation of KK
excitations impossible at the Tevatron collider. (In certain scenarios,
it would be possible to test at Tevatron
compactification scales up to 500 GeV \cite{cmn}).
Second, UED models
provide a natural candidate for dark matter in the lightest KK excitation (LKP).
However, agreement with the experimentally determined value of the current dark
matter density requires that the mass of the LKP be in the 500 - 800 GeV
range \cite{Servant02} (although this constraint might be relaxed if
coannihilation effects are important \cite{kribs05}),
 which almost brings it into
conflict with the precision constraints results.

For these reasons, we consider worthwhile to revisit in this paper
the analysis performed in \cite{flacke}. Our results contradicts 
the conclusions of \cite{flacke}, while being in rough agreement 
with \cite{a_yee}). The outline of the paper is as follows.
In section II we will review the definitions of the parameters relevant
to our analysis, as well as the the SM contributions  and
the experimental constraints on these parameters. In section III, we consider
the magnitude of new physics contributions, and derive the constraints on
the compactification scale. We find a result in  agreement with
\cite{a_yee}; that is, for low Higgs mass (115 GeV), we find a slightly
stronger constraint
$M \gtrsim 600$ GeV (for a top mass value $m_t = 173$ GeV)
which gets weaker with increasing Higgs mass, so
that values of the compactification scale as low as 300 GeV are still allowed
for heavy SM Higgs. We end with conclusions.

\section{Observables and Standard Model predictions}

We start our discussion by defining the relevant parameters and
reviewing the experimental constraints on these
obtained
from measurements at the LEP collider. We will use the epsilon parameter
analysis introduced in \cite{altarelli}.
The data on the basic physical observables
measured at the Z pole
(the ratio of the gauge boson masses, the $Z$ boson decay widths
and the forward/backward asymmetries)
can be interpreted as constraints on three parameters $\eps_1,  \eps_2,
\eps_3$. These parameters, while being zero at tree level, get
contributions from loop diagrams, due to SM as well as to new physics.
Note that the light SM fermions (as well as photon) loops
contribute to vertex corrections
as well as box diagrams, while heavy particles
(like the Higgs boson or the top quark) contribute mainly to the gauge
bosons self energies.  One
can express the contribution of new physics to the $\eps_i$
parameters in terms of the vacuum polarization
functions of the gauge bosons $\Pi_{XY}$ (with $X,Y$ standing for $B$ and
$W_i$) and their derivatives at zero momentum transfer:
\bea{def_eps}
\epsilon_1 &=& \epsilon_{1,SM}+\hat{T}-W+2X\frac{\sin\theta_W}{\cos\theta_W}
-Y\frac{\sin^2\theta_W}{\cos^2\theta_W}\nonumber\\
\epsilon_2 &=& \epsilon_{2,SM}+\hat{U}-W+2 X\frac{\sin\theta_W}{\cos\theta_W} - V \\
\epsilon_3 &=& \epsilon_{3,SM}+\hat{S}-W+\frac{X}{\sin\theta_W\cos\theta_W }-Y .\nonumber
\eea
with the SM contributions encapsulated in $\eps_{i,SM}$, and only new
physics contributing to the parameters
\bea{STdefs}
\hat{T} & = & \frac{1}{m_W^2}\left(\Pi_{W_3W_3}(0)-\Pi_{W^+W^-}(0)\right)\nonumber\\
\hat{S} & = & \frac{g}{g'}\Pi'_{W_3B}(0)\nonumber\\
\hat{U} & = & \Pi'_{W^+W^-}(0)-\Pi'_{W_3W_3}(0)\nonumber\\
X & = & \frac{m_W^2}{2}\Pi''_{W_3B}(0)\\
Y & = & \frac{m_W^2}{2}\Pi''_{BB}(0)\nonumber\\
W & = & \frac{m_W^2}{2}\Pi''_{W_3W_3}(0)\nonumber \\
V & = &  \frac{m_W^2}{2} \left(\Pi''_{W_3W_3}(0)-\Pi''_{W^+W^-}(0)\right)
 \nonumber
\eea
We use here the notations in \cite{barbieri}; the $\hat{S}, \hat{T},  \hat{U}$
parameters are related to the usual $S,T, U$ defined in \cite{Peskin_T} by
$S = 4 \sin \theta_W^2 \hat{S}/\alpha $, $T= \hat{T}/\alpha $,
$U = -4 \sin \theta_W^2 \hat{U}/\alpha $.

For the experimental constraints on the parameters $\eps_i$ we use the
following values \cite{barbieri}:
\begin{equation}\label{eps_val}
    \begin{array} {ccc}
        \begin{array}{c}
            \epsilon_1=+(5.0\pm1.1)10^{-3}\\
            \epsilon_2=-(8.8\pm1.2)10^{-3}\\
            \epsilon_3=+(4.8\pm1.0)10^{-3}\\
        \end{array} &
        \begin{array}{c}
                                          \\
            \mbox{with correlation matrix}\\
                                          \\
        \end{array} &
                 \rho= \left(\begin{array}{ccc}
              1 & 0.66 & 0.88\\
              0.66 & 1 & 0.46\\
              0.88 & 0.46 & 1\\
        \end{array}\right) \ ,
    \end{array}
\end{equation}
which ar the same as those used in \cite{flacke}.
In order to translate these into constraints on the $\hat{S}, \hat{T}, \hat{U}
\ldots$ parameters, we need first to evaluate the SM contributions
$\eps_{i,SM}$. The main contributions to these
parameters are due to loops involving the heavy top quark and Higgs
boson. At first order, and keeping only the
leading terms in $m_t, m_H$, one obtains:
\bea{one-loop}
\eps_{1,SM} & \simeq & { 3 G_F m_t^2 \over 8 \pi^2 \sqrt{2} } \ - \
{ 3 G_F m_W^2 \over 4 \pi^2 \sqrt{2} } \tan^2 \te_W \ln {m_H\over m_Z}
\nonumber \\
\eps_{2,SM} & \simeq & -{ G_F m_W^2 \over 2 \pi^2 \sqrt{2} } \ln {m_t\over m_Z}
 \nonumber \\
\eps_{3,SM} & \simeq & {G_F m_W^2 \over 12 \pi^2 \sqrt{2} } \ln {m_H\over m_Z}
\ - \ { 3 G_F m_W^2 \over 4 \pi^2 \sqrt{2} } \ln {m_t\over m_Z}
\eea
Note that only $\eps_{1,SM}$ has quadratic dependence on the top quark mass.
The leading order estimates aquire corrections from vertex and box
diagrams due to light fermions, as
well as two loop contributions to the gauge boson self energies.
The results can  parametrized by the following
expressions \cite{barbieri}:
\bea{SMepsilon}
\epsilon_{1,SM}&=&\left(+6.0-0.86 \ln \frac{m_H}{m_Z}\right)10^{-3}\nonumber\\
\epsilon_{2,SM}&=&\left(-7.5+0.17 \ln \frac{m_H}{m_Z}\right)10^{-3}\\
\epsilon_{3,SM}&=&\left(+5.2+0.54 \ln \frac{m_H}{m_Z}\right)10^{-3},\nonumber
\eea
evaluated at $m_t = 178$ GeV, and where the dependence on $m_H$ is explicit.

We make a short comment on the first parameter $\eps_{1,SM}$. Using the
definitions for $\eps_i$ in \cite{altarelli,barbieri_susy}, one can write
\beq{e1}
\eps_{1,SM} \ = \ \delta \rho  + M_Z^2 F'_{ZZ}
(M_Z^2) - {\del G_F^{V,B} \over G_F } - 4 \del g_A^{V,B} \ ,
\eeq
where for the $A$ gauge boson $F_{AA}(q^2) = (\Pi_{AA}(q^2) - \Pi_{AA}(0))/q^2$,
and $ \del G_F^{V,B}, \del g_A^{V,B}$ are the vertex and box corrections
to the low energy value of the Fermi constant and to the value of the
axial couplings at $M_Z$. The important thing about the above expression is
that the strong dependence of $\eps_{1,SM}$ on the heavy top mass is
confined to the $\del \rho$ parameter, defined as:
\beq{del_rho}
\del \rho \ = \    {\Pi^{SM}_{ZZ}(0) \over M_Z^2 }
-   {\Pi^{SM}_{WW}(0) \over M_W^2} \ .
\eeq
The leading top quark contributions to $\del \rho$ have
been  evaluated up to two-loops; one can write
\beq{rho_exp}
\del \rho  \ = \ \del \rho^{\alpha} +
\del \rho^{\alpha \alpha_s} + \del \rho^{\alpha^2} \ ,
\eeq
with the leading contribution $\del \rho^{\alpha} =  3 x_t$,
$x_t = G_F M_t^2/8\pi^2\sqrt{2}$ , appearing in Eq. (\ref{one-loop}).
The most important correction to the leading term comes from the QCD gluon
loops
\cite{sirlin_rhoqcd}
\beq{rho_qcd}
\del \rho^{\alpha \alpha_s}  \ = \ - 3 x_t {2\over 9}(\pi^2 + 3)
{\alpha_s \over \pi}   \ \approx 3 x_t \delta_{QCD}
\eeq
which amounts to about -11\% of the leading order contribution ($\alpha_s^2$
corrections have also been computed, and they are small). By contrast,
the two-loop order $\alpha^2$ contributions vary from  about
-1.2\% (for a low Higgs mass) to -1.6\% (for a high Higgs mass)
of the leading contribution \cite{degrassi}. We then rewrite
the first equation in (\ref{one-loop}):
\beq{eps_1SM}
\epsilon_{1,SM} \ = \
3 x_t (1 + \delta_{QCD} ) +
\left(-2.86 - 0.86 \ln \frac{m_H}{m_Z}\right)10^{-3}\ ,
\eeq
such that the dependence of the leading terms on the top mass is kept explicit.

\section{Constraints on new physics}

The massive KK excitations contribute to the oblique parameters $S,T,U$ terms
proportional to the mass splittings between the heavy particles at each KK
level. Thus, the top quark excitations will contribute terms of order
$m_t^2/M^2$, the gauge boson excitations will contribute
terms of order $(m_W^2,m_Z^2)/M^2$, while the Higgs excitations will give
contributions $\sim (m_H^2,m_W^2,m_Z^2)/M^2$. The top quark contributions
are moreover enhanced by a term $m_t^2/m_W^2$. We therefore can expect that
for small Higgs mass (of order 100 GeV),
the terms associated with the top quark excitations will
be dominant, while for larger values of the Higgs mass (which could be of
the same order of magnitude as $M$), the terms associated
with the Higgs boson excitations may give significant contribution.

Let us consider first the constraints on the scale of new physics $M$
in the approximation where the subleading terms $X, Y, W, Z$ in Eq.
(\ref{def_eps}) are neglected, and the Higgs mass is small. Moreover,
let us neglect the $\hat{U}$ parameter; as has been argued in previous
analyses (and as it can be seen bellow), its magnitude is negligible.
 Then Eqs.
(\ref{def_eps}),(\ref{eps_val}), evaluated for a value $m_H = $115 GeV,
give the following constraints
on the  $\hat{S}$ and $\hat{T}$ parameters:
\beq{con_1}
\begin{array}{ccc}
            \hat{T} & = &  (0.5 \pm 0.8) \times 10^{-3} \\
            \hat{S} & = &  (-0.01 \pm 0.9) \times 10^{-3} \
\end{array}\ , \hbox{with correlation} \ \rho = 0.86 \ ,
\eeq
(we neglect the theoretical errors on $\eps_{1,SM},\eps_{3,SM})$.
We use here and in the following
a value $m_t = 173 $ GeV, (the latest Tevatron analysis indicates
$m_t = 172.5 \pm 2.3$ GeV \cite{topmass}).
Note also that with $\hat{U}=0$, the SM value for $\eps_{2,SM}$
is about one sigma away from the experimental average. Due to the
correlations between the three parameters, this leads to a displacement of the
mean values for $ \hat{S}, \hat{T}$ from (-0.5, -0.2)
(derived from Eqs. (\ref{def_eps}),(\ref{eps_val})) to the values given in
(\ref{con_1}) above.

For purposes of clarity, we coment briefly on the methodology we use
for the multiparameter fit.
The goodness of the fit is obtained by evaluating the $\chi^2$
function
$$\chi^2 \ = \ \sum_{i,j}(\eps_i - \mu_i)(\si^2)^{-1}_{ij} (\eps_j - \mu_j) ,
$$
with $(\si^2)_{ij}= \si_i \rho_{ij} \si_j$, the $\mu_i$ and $\si_i$ are
the mean experimental values and the errors for the observables
$\eps_i$, and $\rho$ is
the correlation matrix.  For an analysis taking into account only a subset $A$
of observables, with values of the other observables set to
their predicted SM values $\eps_k^0$,
one derives and uses a new function $\chi_A^2$ such that:
$$\chi^2 \ = \ \chi_A^2 + \chi^2_{min} \ \ , \ \
 \chi^2_A \ = \ \sum_{i,j \in A}
(\eps_i - \mu'_i)(\si'^2)^{-1}_{ij} (\eps_j - \mu'_j) \ .
$$ Note that the new mean values and errors $\mu'_i $ and $\si'_i$ are
equal to the old ones only if $\eps_k^0 = \mu_k$, for $k \notin A$. Moreover,
the confidence level limits are defined in terms of the number of parameters
involved in the fit; so, for example, for a two parameter fit, a 90\%CL
limit corresponds to $\chi^2 = 4.61$, while  a 99\%CL
limit corresponds to $\chi^2 = 9.21$ (see \cite{PDG}).

 The leading order contributions of new physics to the
$\hat{S}, \hat{T}, \hat{U}$ parameters are given by
\bea{ST_lo}
\hat{T} & = & {3 g^2 \over 2 (4 \pi)^2} {m_t^2 \over m_W^2 } \left( {2\over 3}
{m_t^2 \over M^2}  \right)\zeta(2) \ + \
{ g^2 s_w^2 \over (4 \pi)^2  c_w^2}
\left( -{5\over 12} {m_H^2 \over M^2} \right)\zeta(2) \nonumber \\
\hat{S} & = & {3 g^2 \over 4 (4 \pi)^2}
\left( {2\over 9} {m_t^2 \over M^2}  \right)\zeta(2) \ + \
{ g^2 \over 4 (4 \pi)^2}
\left( {1\over 6} {m_H^2 \over M^2} \right)\zeta(2) \nonumber \\
\hat{U} & = &  { g^2 s_w^2 \over  (4 \pi)^2} {m_W^2 \over M^2}
\left( {1\over 6}  \zeta(2) - {1\over 15}
{m_H^2 \over M^2} \zeta(4) \right)
\ ,
\eea
where $s_w, c_w$ are the sine and cosine of the Weinberg angle, and
the $\zeta(2) = \sum1/n^2 = \pi^2/6$ factor accounts for the sum
over KK levels in one extra dimension\footnote{The full expressions used for the
gauge boson vacuum polarization functions are given in \cite{a_yee}. We
have independently evaluated the leading parameters, and we find
complete agreement with \cite{a_yee}.}. Note that for two or more extra
dimensions, the sum over KK states becomes divergent, and one has to
parametrize the contributions from higher energies through effective operators.
(However, although such models have interesting physical consequences 
\cite{Dob-2d}, we do not discuss them here). 
Even for one extra dimension, unitarity arguments suggest that one should
restrict the sum to only the first ten KK levels or less \cite{Dicus-uni};
 then the $\zeta(2) \simeq 1.645$ factor should be replaced by 1.55.
Also note that the top quark
contribution to the $\hat{T}$ parameter is a factor of 2 larger than the result
given in the original paper \cite{acd},
 which partly explains the increase in the lower limit on the
compactification scale from the $\sim $ 300 GeV
value given in \cite{acd}  to about 600 GeV in subsequent works.

\begin{figure}[t!] 
\centerline{
   \includegraphics[height=3.in]{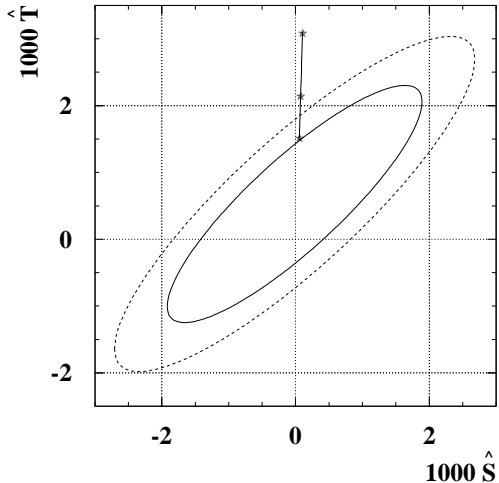}
   }
\caption{The straight line represents the
prediction of the UED model in $\hat{S}, \hat{T}$ plane, for values of
the parameter $M$ from 400 GeV to 600 GeV. The $\star$ symbols correspond
to values $M= 400$ GeV (the upper one), 500 GeV (the middle one) and 600 GeV
(the lower one). The ellipses represent the 90\% CL (solid line) and 99\% CL
(dashed line) constraints on the $\hat{S}, \hat{T}$ parameters
from Eq. (\ref{con_1}).}
\label{st_plot}
\end{figure}

We show in Fig. 1 the predictions of the UED model as a straight solid line
in  the  $(\hat{S}$,  $\hat{T})$ plane,
evaluated for $m_t = 173$ GeV and $m_{H}=115$ GeV. (From the above expressions,
we see that the $\hat{U}$ parameter is suppressed by a factor
$s_w^2 (m_W/m_t)^2$
with respect to $\hat{S}$; hence we can neglect its variation).
The lower end corresponds to a value
$M = 600$ GeV, while the upper end corresponds to a value
$M = 400$ GeV.
The ellipses are the 90\% CL (solid)
and 99\% CL (dashed) contour lines, evaluated according to Eq. (\ref{con_1}).
We see that in this approximation, one can exclude values of $M$ lower
than 550 GeV, for a Higgs mass $m_H = $ 115 GeV. This result is
consistent with the one obtained by Appelquist and Yee  \cite{a_yee},
for the same value of the top quark mass.

One can also use Fig. 1 to estimate what the effect of increasing the Higgs
mass will be. Since the Higgs contribution to the $\hat{T}$ parameter appears
with opposite sign with respect to the top contribution (and with the same
sign for the $\hat{S}$ parameter), the solid line corresponding to the UED
prediction will move downward and to the right. Also, from
Eqs. (\ref{def_eps}),(\ref{eps_val}), one can see that increasing the Higgs
mass will increase the mean value of $\hat{T}$ and decrease the mean value
of $\hat{S}$; hence the ellipses in Fig. 1 will move upward and to the left.
The conclusion is that the constraints on the compactification scale will
soften with increased Higgs boson mass.

The expressions for the subleading parameters in the small Higgs mass
limit are:
\bea{UV_lo}
{V} & = &  { g^2 s_w^2 \over  (4 \pi)^2} {1\over 120}
\left({m_W^2 \over M^2} \right)^2 \zeta(4) \\
{W} & = &  {3 g^2  \over 2 (4 \pi)^2} {m_W^2 \over M^2}
\left( {4\over 15}  \zeta(2) - {1\over 10}
{m_t^2 \over M^2} \zeta(4) \right) \ + \
{ g^2  \over (4 \pi)^2} {m_W^2 \over 60 M^2}
\left( -  \zeta(2) + {1\over 4}
{m_H^2 \over M^2} \zeta(4) \right)
\nonumber \\
{X} & = &  {3 g g'  \over  (4 \pi)^2} {m_W^2 \over 2 M^2}
\left({m_t^2 \over 180 M^2} \right) \zeta(4) \ + \
{ g g'  \over  (4 \pi)^2} {m_W^2 \over 2 M^2}
\left({- m_H^2 + 2 m_W^2 +3 m_Z^2 \over 240 M^2} \right) \zeta(4)
\nonumber \\
{Y} & = &  {3 g'^2  \over 2 (4 \pi)^2} {m_W^2 \over M^2}
\left( -{34\over 135}  \zeta(2) + {77\over 540}
{m_t^2 \over M^2} \zeta(4) \right) \ + \
{ g'^2  \over (4 \pi)^2} {m_W^2 \over 120 M^2}
\left( -  \zeta(2) - {1\over 4}
{m_H^2 \over M^2} \zeta(4) \right)
 . \nonumber
\eea
We have kept in these expressions terms up to $(m_{T,Z,H}/M)^4$.
Note that in this limit, the third family quark loops do not contribute to
the $\hat{U}$ and $V$ parameters. $X$ is also strongly suppressed.
As a consequence,  these three parameters do not contribute significantly
to the constraints on $M$ scale.

The most important subleading parameter is $W$ (the $Y$ parameter
is suppressed by a $(s_w/c_w)^2$ factor with respect
to $W$). Numerically its  magnitude
is about half that of the $\hat{S}$ parameter (at $m_H$ = 115 GeV). As to its
effect on the constraints on the $\hat{T}, \hat{S}$ parameters coming
from Eqs. (\ref{def_eps}), note that it contributes with opposite sign from
$\hat{T}, \hat{S}$ to $\eps_1, \eps_3$.
In consequence, taking $W$ into account would have the
effect of moving a point on the solid line in Fig. 1 towards smaller values
of $\hat{T}, \hat{S}$, that is, roughly parralel to the contour lines
corresponding to a given $\chi^2$. Then one would expect that it will not
have a significant impact on the constraints obtained in the approximation
when only the leading oblique parameters are taken into account.

\begin{figure}[t!] 
\centerline{
   \includegraphics[height=3.in]{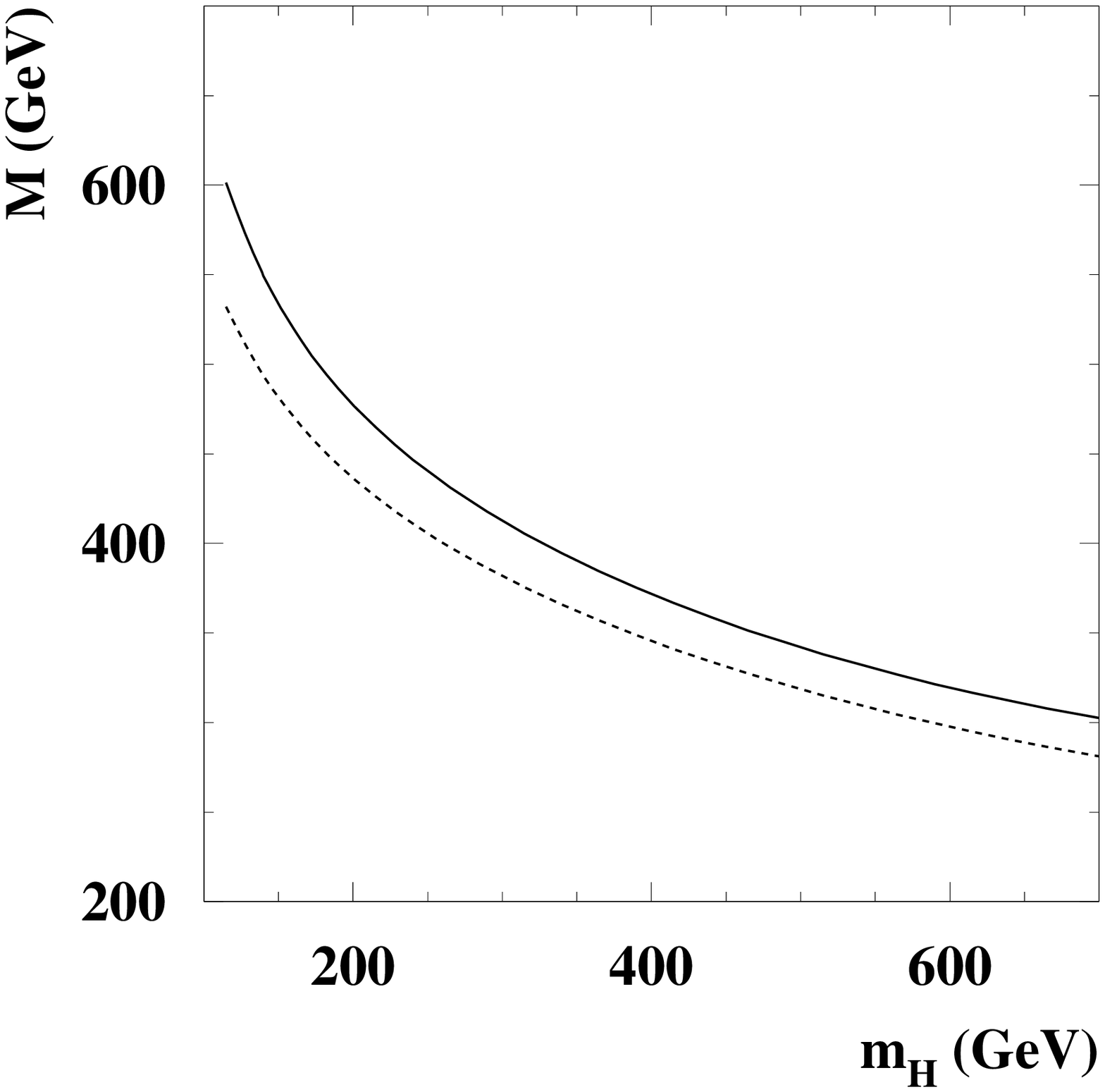}
   \includegraphics[height=3.in]{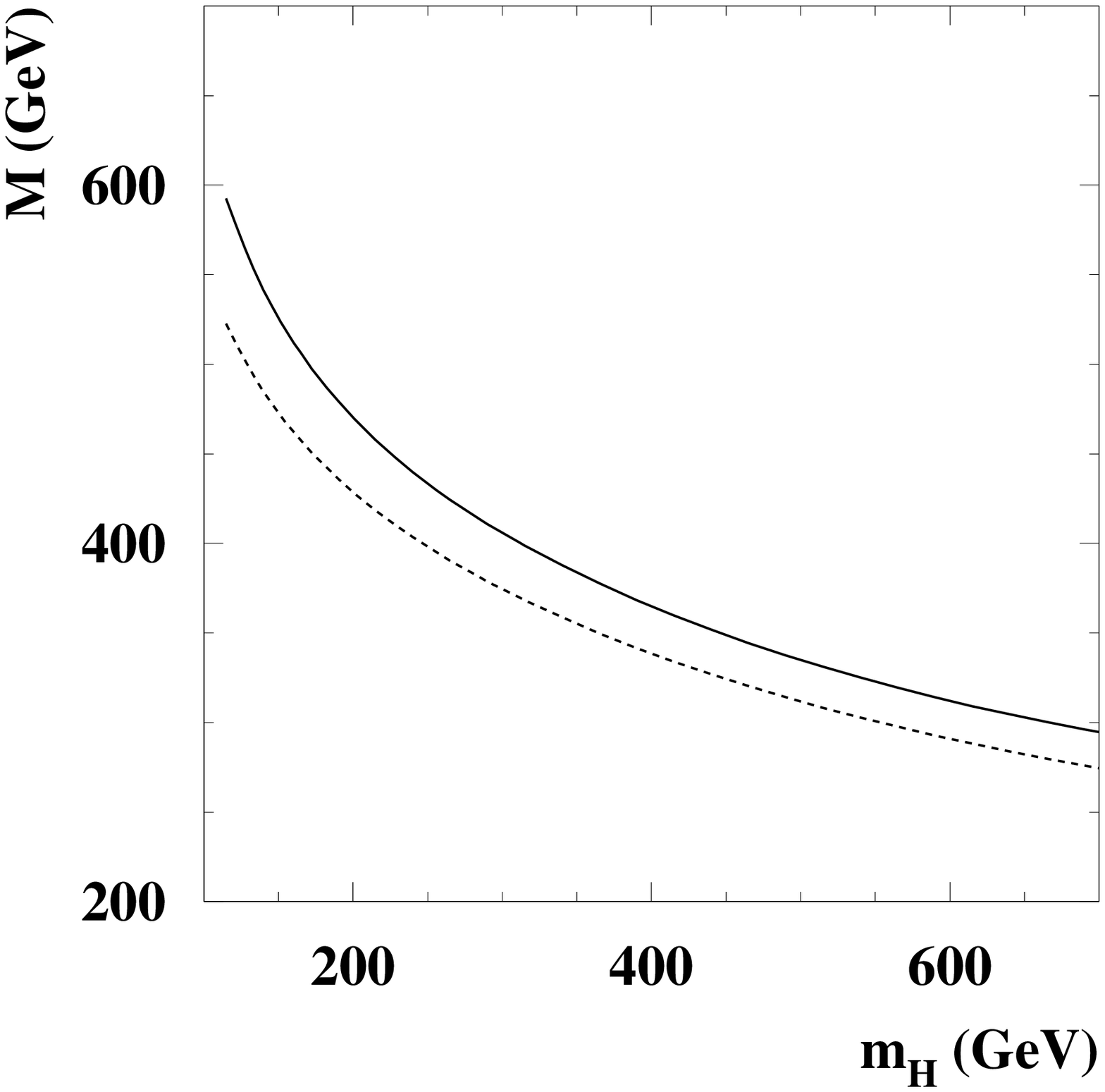}
   }
\caption{Constraint on inverse compactification
scale $M$ as a function of $m_H$, for a top mass value $m_t = 173$ GeV.
Left plot: results obtained using just the $\hat{S}, \hat{T}$
parameters (solid line corresponds to 90\% CL, dashed line to 99\% CL).
Right plot: results obtained from constraints
on the three $\eps_i$ parameters,
 including the subleading contributions from $X, Y, V$ and $W$.}
\label{MH_plot}
\end{figure}

We show in Fig. 2 the resulting constraints (at 90\% CL and 99\%CL) on
the compactification scale $M^{-1}$ as a function of the Higgs mass. The
left plot is obtained using the ST analysis from Eq. (\ref{con_1}) (the
$\hat{U}, X,Y,...$ parameters are set to zero). The result is similar to
the one reported in \cite{a_yee}; the somewhat tighter constraints we obtain
may be attributed to the modifications to the SM constraints on the $S,T$
parameters (one notes that the preffered valued for $S,T$ have moved
significantly towards negative values for the 2004 Particle Data Group
numbers as
compared with the 2002 PDG numbers, even after taking into account
variations due to modifications in the top mass).
Also, we found that the effect of
taking into account the two loop electroweak corrections to $\del \rho$
is not significant. This can be understood by noting the fact that the
term $\del \rho^{\alpha^2}$ is of order $10^{-4}$ itself, and moreover that its
dependence on the Higgs mass is not very strong  (the parametrization in
\cite{degrassi} indicates a roughly logarithmic dependence). Therefore,
the two loop term cannot affect the cancellation between the one loop Higgs
contribution to $\eps_{1,SM}$ and the KK contribution to
the $T$ parameter, contrary to the argument given in \cite{flacke}.

In the right panel in Figure 2 we show the constraints on the compactification
scale obtained by a $\chi^2$ fit for the three $\eps_i$ parameters in Eq.
(\ref{def_eps}),
and taking into account the contributions from the $\hat{U}$ as well as the
subleading  $X,Y,...$
parameters. (The 90\% CL limit corresponds to a $\chi^2 = 6.25$, while
99\% CL limit corresponds to a $\chi^2 = 11.34$).
As expected, we find very small changes (the constraints are somewhat
weakened at small Higgs mass and slightly tightened for large Higgs mass)
compared with the result of the $S,T$ analysis. 

Independent  constraints on the $X,Y,W$ parameters can be obtained by
studying $e^+ e^-$ collisions at energies higher than the $Z$ pole 
(the LEP-II data, see, for example, \cite{lepII}). With the parametrization 
of  new physics we use, this data gives the following constraints 
\cite{barbieri} (independent
of the top and Higgs masses):
\beq{Xvals}
    \begin{array} {ccc}
        \begin{array}{c}
            X=(-2.3\pm3.5)10^{-3}\\
            Y=(+4.2\pm4.9)10^{-3}\\
            W=(-2.7\pm2.0)10^{-3}\\
        \end{array} &
        \begin{array}{c}
                                          \\
            \mbox{with correlations}\\
                                          \\
        \end{array} &
                 \rho= \left(\begin{array}{ccc}
              1 & -0.96 & +0.84\\
              -0.96 & 1 & -0.92\\
              +0.84 & -0.92 & 1\\
        \end{array}\right).
    \end{array}
\eeq
Since they come from different measurements, there are no correlations between
these constraints and the ones on the $\eps_i$ parameters (\ref{eps_val}).
It is then straightforward to see that taking (\ref{Xvals}) into account
has a minimal effect on the results discussed above, since the errors on
these parameters are significantly larger than the errors on the $\eps_i$'s,
and the mean values are consistent with very small (or zero) $X,Y,W$.

\begin{figure}[t!] 
\centerline{
   \includegraphics[height=3.in]{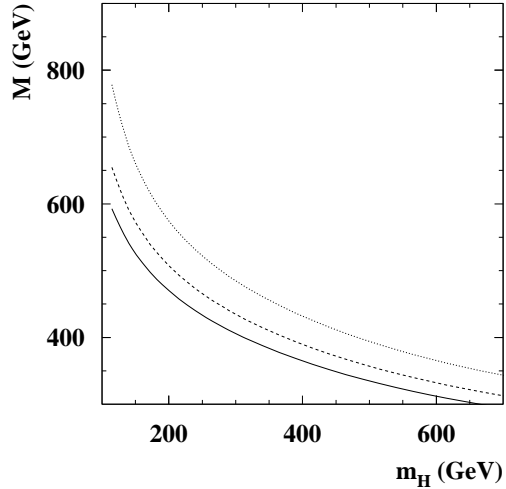}
   }
\caption{Constraint on inverse compactification
scale $M$ as a function of $m_H$, for top mass values $m_t = 173$ GeV
(solid line) 175 GeV (dashed line) and 178 GeV (dotted line).
Results are obtained by
 including the contributions from the subleading terms.}
\label{Mtop_plot}
\end{figure}

Finally, we show in Fig. 3 the 90\% CL limits on the compactification
scale for three different values of $m_t$. Note that these constraints tighten
considerably with increased $m_t$. This is due to the increase in the
magnitude of the UED
prediction for the $\hat{T}$ parameter (which behaves like $m_t^4$), as well
as to the change in the SM value for the $\eps_{1,SM}$ parameter
(\ref{eps_1SM}), which increase with $m_t$, thus pushing the experimental
constraints on $\hat{T}$ to lower values. Thus, for a
top mass $m_t = 178$ GeV, the low Higgs mass constraint
on the compactification scale increases to $M \gtrsim 800$ GeV, which
is in rough agreement with the result of \cite{flacke} at $m_H= 115$ GeV.
However, unlike \cite{flacke}, we find that for increased Higgs mass,
the constraint on the compactification scale weakens, such as that for Higgs
mass of order 600 GeV, $M$ can be as low as $\sim$ 400 GeV.
\section{Conclusions}

We have recalculated  the constraints on UED models arising from
precision electroweak data, taking into account 
two loop electroweak correction
involving the top quark and Higgs boson to the SM parameters, as well as
subleading terms due to new physics.  We consider
the case with one UED.  In this case we found that taking  the SM
Higgs mass 115 GeV and  top quark mass 173 GeV, the lower bound 
on the inverse
compactification scale should be  $M \gtrsim 600$ GeV, in agreement
with previous results. Also we find that this bound is weakened
for increased values of the Higgs mass, in agreement with the results 
of \cite{a_yee}, but in contradiction with those in \cite{flacke}. As a
consequence, the values for $1/R$ preferred by the
models explaining dark matter as stable KK excitations
are still compatible with the electroweak precision constraints.

We also find  that the constraint on
the  compactification scale depends strongly on the top quark mass.
Part of this dependence is due to the $m_t^4$ behaviour of the T parameter;
an equally important part is due to the $m_t^2$ dependence of the SM
$\delta \rho$ parameter. Thus, if the the preferred value for the top mass 
increases by around 3\%, one obtains a roughly 30\% strengthening of the
bound on the compactification scale $1/R$. However, 
even with a value of 178 GeV
for the top mass, one can have a compactification scale as low as 400 GeV 
(with a large Higgs mass), thus 
potentially allowing for the observation of KK excitations at the Tevatron 
Run II.

\section{Acknowledgments}

We would like to thank K. Agashe,  B. Dobrescu, C.N. Leung and Q. Shafi 
for useful
discussions. This work is supported in part by the US Department
of Energy grants number DE-FG02-85ER40231 (C.M.)  and
DE-FG02-84ER40163 (I.G.).

\end{document}